\documentclass[aps, twocolumn, floatfix, amsmath]{revtex4-1}

\usepackage[utf8]{inputenc}
\usepackage{physics}
\usepackage{graphicx}
\usepackage{amsmath,amssymb}
\usepackage[colorlinks,allcolors=blue]{hyperref}
\usepackage{natbib}

\begin{document}
\title{Condensation to a strongly correlated dark fluid of two dimensional dipolar excitons}

\author{Yotam Mazuz-Harpaz}
\author{Kobi Cohen}
\author{Ronen Rapaport}
\affiliation{Racah Institute of Physics, The Hebrew University of Jerusalem, Jerusalem 9190401, Israel.}

\date{\today}

\begin{abstract}
Recently we reported on the condensation of cold, electrostatically trapped dipolar excitons in GaAs bilayer heterostructure into a new, dense and dark collective phase. Here we analyze and discuss in detail the experimental findings and the emerging evident properties of this collective liquid-like phase. We show that the phase transition is characterized by a sharp increase of the number of non-emitting dipoles, by a clear contraction of the fluid spatial extent into the bottom of the parabolic-like trap, and by spectral narrowing. We extract the total density of the condensed phase which we find to be consistent with the expected density regime of a quantum liquid. We show that there are clear critical temperature and excitation power onsets for the phase transition and that as the power further increases above the critical power, the strong darkening is reduced down until no clear darkening is observed. At this point another transition appears which we interpret as a transition to a strongly repulsive yet correlated $e$-$h$ plasma.    
Based on the experimental findings, we suggest that the physical mechanism that may be responsible for the transition is a dynamical final-state stimulation of the dipolar excitons to their dark spin states, which have a long lifetime and thus support the observed sharp increase in density. Further experiments and modeling will hopefully be able to unambiguously identify the physical mechanism behind these recent observations.  
\end{abstract}

\maketitle

\section{Introduction}
The understanding of the collective effects of ultra-cold, quantum degenerate bosonic and (more lately) fermionic gases has advanced significantly in the last two decades. This, mostly due to major advances in systems of cold atomic species, but more recently also in condensed matter systems, where cold gases of electronic excitations have been realized \cite{snoke_spontaneous_2002,snoke_coherence_2010,deveaud_exciton-polariton_2015}. One unique class of bosonic many-body systems is that of dipolar fluids, where the combination of Bose-Einstein quantum statistics and the long ranged dipole-dipole interaction gives rise to a rich and unique fundamental physics. Unlike the case of short range interaction, where particles are essentially free except during instantaneous collision events, dipolar particles interact with each other over long distances. In this sense, dipolar fluids are much more correlated than weakly interacting fluids. The dipolar interaction-induced correlations between particles can lead to very interesting effects such as pattern formation and instabilities even for purely classical particles \cite{cowley_interfacial_1967}. Even more exotic are cases where the classical correlations compete with quantum mechanical effects. This usually happens whenever the quantum kinetic energy of each particle (due to the momentum-position uncertainty principle) is of the same order of magnitude as the typical interaction energy between particles. This makes the center-of-mass position of particles in a dense interacting fluid uncertain, leading to a significant wavefunction overlap between particles and signifying the onset of collective Bose-Einstein quantum effects. One well known system where such competition prevails is cold $^4$He where this interplay leads to the condensation of the liquid into a superfluid state at low enough temperatures. In $^4$He the long-range part of the interaction is attractive, which leads to the formation of a stable liquid at low enough temperatures and to the transition into a quantum liquid at lower temperatures. These long range interactions are of Van Der Waals type and therefore they fall quickly with the distance between particles in the fluid. As a result, the gas phase of $^4$He is only weakly correlated. It is far less obvious to figure out what are the correlation effects of a fluid of dipoles, where the interactions decay slower with the particle distance. In particular, it is interesting to understand the stable phases of a fluid of particles where this long range interaction is purely repulsive. This is true both in the classical regime where the dominant kinetic energy is that of thermal motion and in the quantum regime where the dominant kinetic term becomes the zero-point motion of the particles. From the experimentalist point of view, it is therefore a worthy challenge to realize and study such dipolar fluid in various systems. 

Indeed, such dipolar many-body systems have been realized and studied in semiconductor quantum-well (QW) heterostructures, where indirect excitons are optically excited, for several decades now \cite{perry_magnetooptical_1990,westgaard_optical_1992,kawai_doublet_1985,kash_thermodynamics_1992,chen_effect_1987,charbonneau_transformation_1988,butov_anomalous_1996,butov_direct_1995,bayer_direct_1996,andrews_stark_1988,fukuzawa_phase_1990,fukuzawa_possibility_1990,alexandrou_electric-field_1990,butov_cold_2007,rapaport_experimental_2007}, and more recently they were also demonstrated in atomic physics \cite{lahaye_physics_2009,baranov_condensed_2012}, where some striking observations related to quantum dipolar correlations and condensation have already been observed \cite{santos_bose-einstein_2000,buchler_strongly_2007,baranov_ultracold_2002,jin_polar_2011,lu_strongly_2011,yan_observation_2013,gaj_molecular_2014,griesmaier_bose-einstein_2005,aikawa_bose-einstein_2012,lahaye_strong_2007,kadau_observing_2016,ferrier-barbut_observation_2016,chomaz_quantum-fluctuation-driven_2016}.
In the last few years the field has further expanded to polaritonic systems \cite{christmann_oriented_2011, cristofolini_coupling_2012, rosenberg_electrically_2016}, to bilayer two-dimensional transition metal DiChalcagonide systems \cite{rivera_observation_2015} and to bilayer graphene \cite{li_negative_2016,li_excitonic_2016}. All these realizations of cold dipolar fluids have already led to new and exciting, sometimes unexpected, observations. These observations  are unique to such correlated fluidic systems and there is still a gap to bridge between the multitude of experimental reports and a consistent theoretical framework. 

In this paper we focus on a system of two dimensional (2D) indirect excitons (IXs) in GaAs double quantum well (DQW) heterostructures: a unique system of 2D boson-like dipolar quasi-particles which are dynamically excited using light and that have an internal spin degrees of freedom which determines their decay dynamics.  In recent years the experimental progress has allowed a consistent study of cold IX fluids in a wide range of densities. The striking observations clearly point to an intricate many-body quantum effects that are yet to be fully understood. Despite the currently unanswered questions, this progress has already opened a window into the complex physics of interacting dipolar fluids which are coupled to light in a non-trivial manner.

\begin{figure}
\centering
\includegraphics[width=0.5\textwidth]{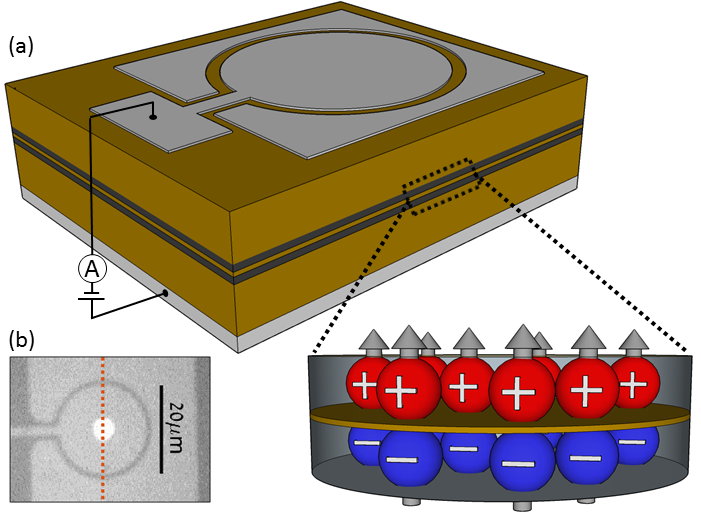}
\caption{\label{fig:sample} (a) An illustration of IXs in an electrostatic trap in a DQW sample. Excitons are created by laser excitation of the trap region. An electric voltage between the conducting substrate and the metallic top gate polarizes the excitons which are confined under the gate. (b) A microscope image of the IX trap used in the experiment described in Sec. ~\ref{sec:experimets}. It is surrounded by a guard-gate, with the bright laser spot in its center.}
\end{figure}

\section{A short review of recent theoretical and experimental progress}
\subsection{Dipolar excitons in GaAs double quantum wells}
An indirect exciton is a Coulomb-bound yet spatially-separated pair of an electron ($e$) and a hole ($h$), typically created optically by laser excitation in an electrically-biased DQW heterostructure \cite{rapaport_experimental_2007, butov_cold_2007}, as illustrated in Fig ~\ref{fig:sample}. Due to this $e$-$h$ spatial separation, IXs carry large electric dipole moments. These are define as $ed/\sqrt{\varepsilon}$ where $d$ is the separation between the $e$ and the $h$, and $\varepsilon$ is the background dielectric constant. Typical dipole lengths for IXs are $d/\sqrt{\varepsilon}\sim 3\mbox{-}6$nm, which are much larger than in their atomic counterparts. The spatial charge separation also leads to long IX radiative lifetimes, typically in the range of $100\mbox{-}1000ns$. As that lifetime is much longer than the thermalization time with the lattice \cite{rapaport_analysis_2006}, IX fluids may reach a quasi thermal equilibrium with it. Additionally, as the effective mass of an exciton is just a fraction of the free electron mass $m_0$, the thermal De-Broglie wavelength is very large and quantum statistics is expected to become important at temperatures as high as several degrees Kelvin \cite{butov_towards_2002,lozovik_strong_2007,rapaport_experimental_2007}. 
In the last two dacades, the interest in cold dipolar fluids of excitons in GaAs-based structures has been growing consistently. The developement of new techniques for trapping and manipulating IXs \cite{snoke_trapping_2005,butov_macroscopically_2002,snoke_long-range_2002,rapaport_electrostatic_2005,hammack_trapping_2006,hammack_excitons_2006,voros_trapping_2006,chen_artificial_2006,schinner_confinement_2013,andreakou_optically_2014,gartner_drift_2006,high_control_2008,leonard_transport_2012,rudolph_long-range_2007,vogele_density_2009,winbow_electrostatic_2011,alloing_optically_2013,gartner_micropatterned_2007,kowalik-seidl_tunable_2012,kuznetsova_control_2010,schinner_electrostatically_2011,violante_dynamics_2014,butov_magneto_2000,butov_observation_2001,butov_spatially_2000,repp_confocal_2014}, allows a better control of excitonic fluids, and quantitative analysis of experiments. This led to a significant progress in the field, and to recent experimental observations of several very intriguing phenomena. While these are not yet fully understood, they indeed signify the complexity and richness of the system, and the importance of the interactions and correlations in the formation of the collective states of dipolar quantum fluids. 

\subsection{Particle correlations in a dipolar exciton gas}
Several theoretical works have shown that due to the combination of the properties mentioned above, the interplay between the interaction-induced many-body correlations and quantum statistics is expected to be very pronounced in IX fluids at very feasible temperature and density ranges (see e.g. \cite{lozovik_strong_2007,schindler_analysis_2008,laikhtman_correlations_2009,laikhtman_exciton_2009}). In fact, strong particle correlations should exist under almost every realistic experimental conditions. At high temperatures and low enough densities strong pair correlations in a classical gas phase are expected, where excitons tend to avoid each other due to their mutual dipolar repulsion. This repulsion is balanced by the thermal kinetic energy of the dipoles in the gas, resulting in a typical average scattering distance $r_0$ between the dipoles given by $E_{dd}(r_0)\sim k_BT$, where $E_{dd}(r)\simeq e^2d^2/\varepsilon r^3$ is the classical interaction between a pair of dipoles. This balance leads to an effective depletion circle which forms around each IX, whose radius $r_0(T)$ increases as the temperature is lowered. Such pair correlations should reduce the average repulsive interactions in the IX gas from the mean field value \cite{schindler_analysis_2008,laikhtman_correlations_2009,laikhtman_exciton_2009} and they should become stronger as the temperature is lowered and the thermal energy of the IXs is reduced. The thermal fluctuations around this average value of interaction energy should also decrease with decreasing temperature. The effect of the increased correlations is therefore expected to be experimentally manifested by the reduction of both the measured repulsive interaction energy of recombining IXs in an IX gas with a fixed density, as well as by the reduction of their linewidth \cite{laikhtman_correlations_2009}. 
A clear experimental evidence for such an effect was reported by \citet{shilo_particle_2013} and later on by other several works \cite{schinner_many-body_2013,remeika_measurement_2015,cohen_dark_2016} for a macroscopic number of IXs, and also for the case of only few IXs in a tiny trap \cite{schinner_confinement_2013}. This effect will be also briefly discussed later on in this paper. These observations have proven that indeed a fluid of IXs is fundamentally different from a weakly interacting gas of either atoms \cite{pitaevskii_boseeinstein_2003}, excitons, or exciton-polaritons \cite{deveaud_exciton-polariton_2015}.
Another interesting low density correlated gas regime is expected to occure when a very dilute gas is cooled down to a temperature where the thermal De-Brogli wavelength of the excitons $\lambda_{db}(T)$ becomes larger than $r_0$. At this point the dipolar scattering between excitons should be described quantum-mechanically. This leads to an effective quantum depletion region which is expected to become temperature independent \cite{laikhtman_correlations_2009,laikhtman_exciton_2009}, which in turn leads to temperature independent pair correlations and temperature independent average repulsive interaction energy. Recently we observed a transition from a temperature regime where the repulsive interaction energy per exciton (also called the "blue shift" energy) decreases with decreasing $T$, as expected from a classically correlated gas, to a temperature regime where the blue shift becomes $T$-independent (with a transition between the regimes that goes like $T^2$) \cite{shilo_particle_2013}. This might be an indication to such classical-to-quantum correlation crossover but further experiments are required to verify this point.

\subsection{Multi-particle correlations and collective phases}
Even more exciting and intriguing are the possible multi-particle correlated phases of an IX fluid, which are expected when the fluid becomes cold and dense enough. Multi-particle interactions are expected when the depletion radius $r_0$ is larger than the actual average inter-particle distance $n^{-1/2}$, i.e., when $r_0(T)n^{1/2}\gg1$. The condition for the onset of multi-paticle interactions $r_0(T)n^{1/2}\sim1$ should be compared to the criterion of the onset of quantum degeneracy, $\lambda_{db}(T_0)n^{1/2}\sim1$. At low enough temperatures and low enough densities, where the quantum degeneracy condition is met and the multi-particle interaction condition is not, a significant wavefunction overlap of neighboring IXs is expected, across the depletion region of each particle. In such case the fluid should undergo a BEC transition with a significant condensate fraction yet observable particle correlations \cite{lozovik_strong_2007}. Several experimental works studying cold IX clouds in exciton rings \cite{butov_macroscopically_2002,rapaport_charge_2004,snoke_long-range_2002} reported a fragmantetion of the ring into ordered beads with extended spatial coherence \cite{high_spontaneous_2012, alloing_evidence_2014, gorbunov_large-scale_2006} below a critical temperature. This was interpreted as a signature for the onset of a Bose-Einstein condensation of the IXs on the ring. A similar onset of extended spatial coherence was reported for IXs trapped in an electrostatic trap \cite{high_condensation_2012}. In those works a critical temperature for the onset of the coherence was reported, but no density threshold was clearly identified. This leaves open the question of which type of a phase transition was observed (e.g. was it a first or a second order transition), and the question of the density regime in which coherence appeared, i.e. whether it corresponds to the dilute limit described above. Long range coherence is expected from a dilute gas of bosons in the limit of weak interaction. It is yet to be fully understood whether the experimental observations are indeed in this range. 

Contrary to that, at high fluid densities, where $r_0n^{1/2}\gg1$ occurs at temperatures that are above $T_0$, a correlated, interacting, multi-particle state should appear, expected to resemble a classical liquid state \cite{laikhtman_exciton_2009}. This dipolar liquid state should be characterized by a short range order and reduced fluctuations, and by a significant reduction of the diffusivity. The latter, due to the quenching of the mean-free-path of the excitons resulting from the multi-particle dipolar interactions. A quantitative theory of a classical gas-liquid transition of 2D dipolar indirect excitons, and of the thermal and mechanical properties of such liquid state, if exists, is not yet available as far as we know. Nevertheless, a clear first-order phase transition of a dense fluid of excitons to a new phase that has a set of features of a liquid state was reported recently by \citet{stern_exciton_2014}. In that report as a spatially extended, high density gas of IXs was cooled down, a new phase emerged to coexist with the initial gas phase, with clear spatial separation between them and a clear phase boundary. The new phase had a lower interaction energy and a smaller, more symmetric linewidth than the gas phase, indicating stronger particle correlations and strongly reduced fluctuations, respectively. A significant reduction in diffusivity was also reported. These properties are a strong signature of a liquid state. A clear critical temperature ($T_c=4.7$K) and a critical excitation power were also observed, with a sharp onset of the liquid state beyond those critical values. That fact, together with the coexistance of the liquid and gas phases, points toward a first order phase transition. However, it is still not completely clear if there was a significant discontinuity in density between the two phases, as expected from a first order phase transition. Notably, because interferometric measurements of the liquid state did not show any extended spatial coherence beyond the diffraction limit, the authors have identified this as an indication that the liquid state is classical, following the prediction by Laikhtman and Rapaport \cite{laikhtman_correlations_2009}.  Interestingly, it was noted in \cite{laikhtman_exciton_2009,laikhtman_correlations_2009} that the onset of a classical liquid state might suppress the onset of quantum degeneracy, with the transition to the quantum liquid state being pushed down to lower temperatures.  This is due to the significant localization of particles in the liquid state that prevents good overlap of the wavefunctions of adjacent particles. A more quantitative prediction of the onset of quantum liquidity of an IX gas through a Berezinskii–Kosterlitz–Thouless transition was performed using numerical Monte-Carlo simulations in Ref. \cite{lozovik_strong_2007,lozovik_superfluidity_2007}. There, an opposite effect was predicted, as it was calculated that the transition temperature to the quantum liquid state grows with increasing dipole density and with increasing particle localization. 

The multi-particle states discussed above, quite generic to a 2D dipolar fluids of bosons, do not however take into account two major inter-related ingredients in every IX fluid in GaAs structures. The first is that IXs are optically created by light and have a finite lifetime. It is finite since they can recombine either via optical emission of photons or by $e$ and $h$ tunneling through the opposite QW barriers towards the two contacts, creating electrical current through the sample. The typical recombination lifetimes in both channels could be very different, with radiative lifetimes of excitons being usually much shorter than tunneling times, and both strongly depend on various experimental parameters such as the applied bias, the temperature, and the density \cite{shilo_particle_2013,mazuz-harpaz_radiative_2016}. This optical generation and the two recombination channels result in a fluid of IXs without a conserved number of particles, and this should have a major impact on the formation dynamics of any collective state. 

The second missing ingredient in the theoretical picture above is the unique internal spin structure of excitons in GaAs heterostructures, which should also have a major role in the formation of their collective dynamics. The total spin projection of heavy-hole excitons is $\pm1$ or $\pm2$, corresponding to two radiative ("bright") states and two non-radiative ("dark") states. In GaAs, where photoluminescence is the dominant depletion mechanism of excitons, this naturally means that the intrinsic lifetime of the dark excitons is much longer than that of the bright ones. However, under normal circumstances, the very fast $h$ spin flip in the IX leads to a full thermalization between the bright and dark states. Since in typical DQW structures all four spin states are also almost degenerate in energy (at all realistically achievable temperatures), it is assumed that the dark and bright populations are equal, and that they posses a single effective lifetime, mostly determined by the radiative lifetime of the bright excitons (see e.g., the supplamentary information of Refs. \cite{shilo_particle_2013,cohen_dark_2016} and \cite{mazuz-harpaz_radiative_2016} for discussion on the IX lifetime). However, due to spin-dependent exchange interaction between the electrons and holes in an IX, the dark excitons are actually slightly lower in energy than the bright excitons \cite{maialle_exciton_1993,glasberg_exciton_1999}. As long as no quantum collective effects are in play, this energy splitting is negligible. However, if a Bose Einstein condensation can occur, ideal bosons macroscopically condense to the lowest energy state regardless of how small the gap from excited states is. It was therefore predicted a few years ago that the ground state of a dilute dipolar exciton fluid is in fact dark and that a macroscopic phase transition to a dark condensate should occur \cite{combescot_bose-einstein_2007}. This work however neglected the interactions between IXs and thus a later work from the same authors considered the case where pairwise short range interactions are introduced. It was shown that these interactions lead to a density threshold for the dark condensate, above which the condensate should become ”gray”, as the interactions mix the bright and dark spin states. The bright component of the condensate was predicted to grow linearily with the density above the threshold density until it reaches an equality with the dark part at high densities \cite{combescot_``gray_2012}. The theory did not calculate what should be the threshold dark density for realistic IX systems, nor it took into account any correlations that arise from the spatially extended dipolar interactions. Additionally, as with the previous theories described above, the IXs were considered to be in full thermal equilibrium and to have a well defined density. Thus, no dynamical treatment was given for the case of finite lifetime IXs under steady state excitation, where the density is not conserved but it is rather determined from the internal dynamics of the system and the external pumping. In that aspect the theoretical picture is still incomplete, as no published theory we are aware of has been put forward taking into account both dipolar interactions, the internal spin structure, and the dynamics of the system under an external particle pumping and internal particle losses that in turn depend  on the relative occupation of the different spin states. We believe that in light of the results described below, such a treatment is essential in order to describe the origin and nature of the ground state of the IXs fluid. This is specifically important because any decoupling between the densities of the dark and bright IXs should obviously have an effect on the dynamics of decay and formation of the fluid, due to the different lifetimes of the dark and bright IXs. 

Experimentally, in \cite{shilo_particle_2013} we reported the first observation of anomalous darkening below a critical temperature of $T\cong 2.6K$, where a sharp increase of the density of an IX fluid in a flat-bottom wide trap occurs, indicated by an increase of luminescence energy blueshift, but with a corresponding decrease of the luminescence intensity. This was interpreted as an evidence for a possible condensation of a macroscopic number of excitons into a dark state. The observation of darkening was later also reported in IX rings \cite{alloing_evidence_2014}, and some darkening was also observed in the liquid state reported in \cite{stern_exciton_2014}. While these works gave indications that a darkening transition does occur, the full experimental picture was still missing. In a recent work we reported on a comprehensive, quantitative experiments of an IX fluid in a quasi-parabolic trap \cite{cohen_dark_2016}. In that work, a very clear phase transition was identified, in which above a critical excitation power and below a critical temperature, a gas of IXs condensed to a dense, compact phase at the center of the trap, initially made of mostly dark particles. This phase is characterized by a spectral narrowing and a sharp increas of the total density, as in a highly correlated liquid. As the excitation power further increased the IXs became less dark yet still maintained their other properties, such as the nearly closely packed density and the almost fixed spatial extent and spectral linewidth. At high enough exitation power the bright and dark densities approximately equated, and another sharp transition was observed, where the density further increased and a large spatial expansion was observed. In the following we summarize the observations and analysis of the experiments reported in that work \cite{cohen_dark_2016}.

\section{Experiments with a dipolar exciton fluid in a quasi-parabolic electrostatic trap} \label{sec:experimets}

\begin{figure*}
\centering
\includegraphics[width=0.95\textwidth]{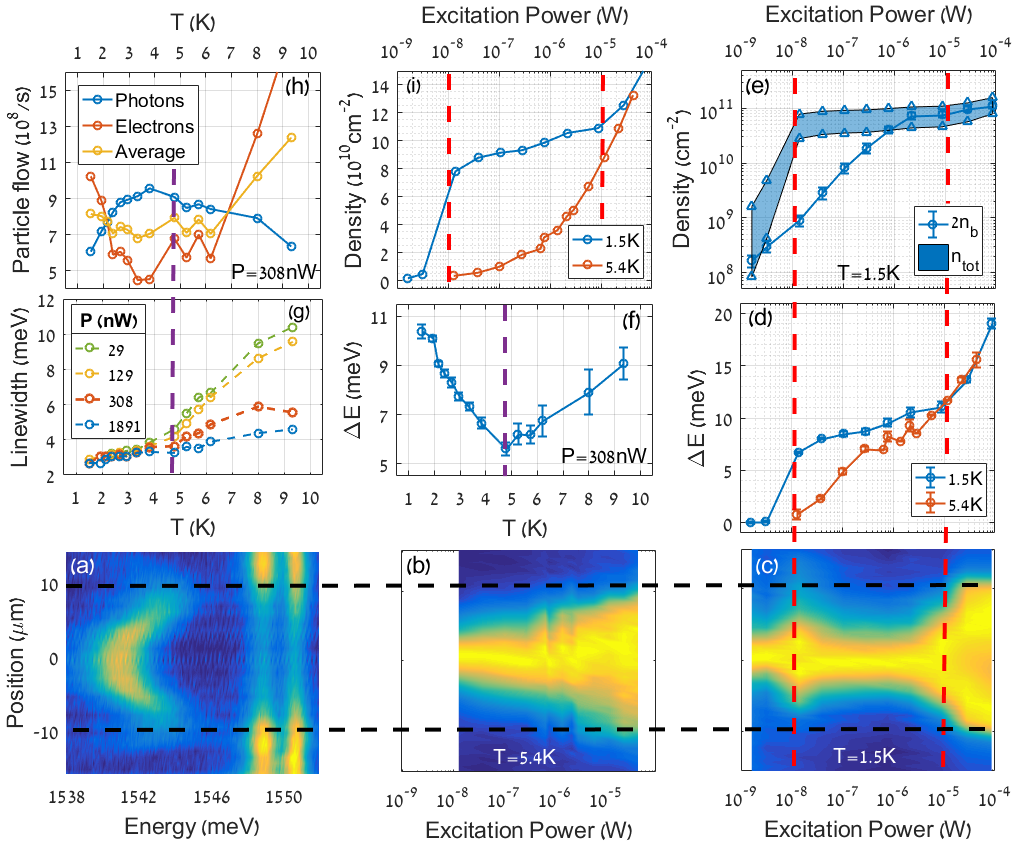}
\caption{
\label{fig:summary} (a) A spatial-spectral profile of the trapping potential, obtained by scanning a weak focused excitation point across the dotted line marked in Fig ~\ref{fig:sample}b. The dashed black lines mark the edges of the trapping gate. The spatial expansion of the IX cloud for different excitation powers for (b) $T=5.4K$ and (c) $T=1.5K$. (d) The interaction energy $\Delta E$ as a function of excitation power for the same two temperatures, below and above $T_c$. The red dashed lines mark the two critical powers $P_{c1}$,$P_{c2}$ separating the three regimes interpreted as the gas, liquid and expanding plasma respectively. (e) The blue shaded region is the extracted total IX density range for $n_{tot}$ as a function of the excitation power at $T=1.5$K. The blue circles denote twice the extracted bright IX density, $2n_b$. (f) The interaction energy $\Delta E$ as a function of $T$ for a constant excitation power of $308nW$. (g) The IXs spectral linewidth as a function of temperature for four exemplary excitation powers. (h) The measured  fluxes of charge carriers in  the tunneling current (red) and of emitted photons (blue). The  orange line is the average of the two fluxes. The purple dashed lines in (f-h) marks $T_c=4.8K$. (i) An estimate of the total IX densities at $T=5.4K$ (red) and  at $T=1.5K$ (blue). The high temperature density is extracted as twice the measured bright density, assuming $n_d\simeq n_b$ above $T_c$ while the low temperature density is extracted from $\Delta E$, according to a liquid model \cite{laikhtman_exciton_2009}, as explained in the text.
}
\end{figure*}

In the experiment reported recently by \citet{cohen_dark_2016}, a trapped IX fluid was studied in an electrostatic trap which was excited at its center by a continuous-wave, non-resonant laser. The trap's shape was circular with a diameter of $20\mu m$, as presented in Fig. ~\ref{fig:sample}b, and its potential shape at low IX density was quasi-parabolic, as shown in Fig. \ref{fig:summary}a. The emission of the trapped bright IXs was monitored as the temperature, the external voltage of the trapping potential and the excitation power were varied. We introduced and employed, for the first time, an experimental technique of constant energy lines (CEL). In this technique the spectral position of the IXs is retained throughout the experiment by adjusting the external voltage such that it compensates for any spectral shifts due to varying experimental conditions (such as temperature and excitation power variations). Since the single-IX properties are mostly determined by the local electrostatic environment of the IX, which also determine its spectral position, maintaining the exciton on a CEL helps in approximately fixing its single-particle properties. This in turn helps tremendously in the analysis of the data, and in extracting the many-body effects. 

With this method, several observations were reported. Firstly, when the trapped fluid was cooled under a constant excitation power, the dipolar interaction energy per particle, measured from its blue shift $\Delta E$, initially decreased, indicating the increase of particle correlations, as was discussed above. Surprisingly, below a temperature of $T_c=4.8$K, the correlated IX gas suddenly displayed a sharp increase of its interaction energy, reaching a constant value $\Delta E_l$ at a lower temperature of $T\cong2$K. This sharp rise in $\Delta E$, plotted for one excitation power in Fig. \ref{fig:summary}f, indicates a sharp increase of the \emph{total} density, as $\Delta E\propto n_{tot}$, where $n_{tot}$ is the total local density of the IX fluid, including both bright and dark particles. A very similar increase to the same final $\Delta E_l$ value was observed for a very wide range of excitation powers, marking that the density increases to a saturated value. To quantify the final density by mapping $\Delta E_l$ into density, requires a knowledge of the correlations between the dipoles \cite{laikhtman_correlations_2009,laikhtman_exciton_2009}. However, one can obtain a range of possible final densities by taking the two limits of possible correlations, namely the totally uncorrelated state (described by a mean field theory) and the highly correlated state (a liquid with short range order) \cite{cohen_dark_2016}. These two limits yield the range of possible saturation densities to be $n_l= 6 \mbox{-}9 \times 10^{10}cm^{-2}$. Since for all values within that range the condition of a multi-particle correlations is well satisfied, namely, $r_0(T=2K)n_l^{1/2}\gg1$, the self consistent mapping between $\Delta E$ and $n$ should be that of a liquid state, which is the highest value in the above range, i.e., $n_l\simeq 9 \times 10^{10}cm^{-2}$. In fact, this value corresponds to approximately close packing of the IX wavefunctions, $\expval{r_l}/a_b\sim 2\mbox{-}3$, where $a_b$ is the IX's Bohr radius. This density buildup was accompanied by a clear reduction of the spectral linewidth, indicating increased particle correlations and the appearence of short range order, as expected from a liquid state. Again, as with the energy shift, the linewidth of the emission was reduced to a saturated value for a wide range of excitation powers below the same critical temperatue $T_c$, as shown in Fig. \ref{fig:summary}g. Notably, It was also accompanied by spatial shrinkage of the IX cloud area towards the center of the trap, indicating either a strong suppression of the IX mobility, a strong suppression of the thermal distribution of the kinetic energy, or the onset of attractive short-range interactions, that perhaps can occur at close packing of IXs \cite{tan_exciton_2005}. All these findings are consistent with the interpretation of a transition into a dense liquid state, having short-range order and a well-defined density which is spontaneously chosen by the system itself. 

After establishing a clear evidence for a sharp transition in temperature, it is also important to show how the transition was manifested as a function of the excitation power, which controls the rate of creation of IXs in the trap. Figs. \ref{fig:summary}c,d present the normalized emission profile of the bright excitons in the trap as a function of the excitation power, for two temperatures, one just above $T_c$, and one well below it, respectively. While for the high temperature, the width of the cloud grows monotonically with increasing power, as expected from a mobile repulsive gas of dipolar particles, the low temperature behaves very differently. The initial expansion at low powers is followed by a significant contraction across a clear critical power $P_{c1}$ to a narrow cloud of constant width. That width is unchanged over a wide range of excitation powers, up to another clear threshold power $P_{c2}$ after which the cloud suddenly expands again, all the way to the boundaries of the trap's gate. The contraction of the cloud above the first critical power is accompanied by a sharp increase in $\Delta E$ to its saturated value $\Delta E_l$, as is seen in Fig. \ref{fig:summary}d. Such a sharp increase is not observed in the high temperature data, above $T_c$. As mentioned above, the blue shift energy can be mapped to a total dipolar particle density, taking into account the possible range of correlations. This is depicted by the blue region in Fig. \ref{fig:summary}e, where a sharp rise to a high density value $n_l$ is observed across the critical power. This value then increases very slowly over a wide range of excitation powers up to the second threshold. Again, all these findings are consistent with a correlated multi-particle state: a liquid. The strong expansion beyond $P_{c2}$ indicates a transition to a new, highly mobile, dense phase, most likely a correlated $e$-$h$ plasma phase, expanding due to the strong in-plane repulsion. The interpreted comparison between the total exciton densities as a function of the excitation power below and above $T_c$ is shown in Fig. \ref{fig:summary}i, where a liquid model is assumed for the low temperature and a mean field model is assumed for the high temperature.

One fundamental question that arises is what mechanism can support such a large increase of density under constant excitation conditions? If the dark and bright IX densities are equal due to thermal equilibrium between them, then at steady state their combined density should be given by $n_{tot}=G\tau_{eff}$. Here, $G$ is the optical generation rate of IXs and $\tau_{eff}$ is the effective lifetime which governs the simultaneous decay of both bright and dark populations. With the CEL method $\tau_{eff}$ should stay constant, at least for a fixed temperature, and thus the density should be proportional to $G$. It is therefore not clear what can cause the sharp density increase. 
A possible solution is that the population equality between the dark and bright states is not maintained below $T_c$ and above $P_{c1}$. An experimental direct and independent measurement of the bright and dark densities, $n_b$ and $n_d$ respectively, is a hard task. However, since $\Delta E$ is a measure of $n_{tot}=n_b+n_d$, and since $n_b$ can be extracted from the collected photon flux, it is possible to deduce also $n_d$ independently. The calibration of the bright density from the measured PL intensity $I$ relies on the relation $I=B\cdot n_b/\tau_b$, where $B$ is a coefficient taking into account many experimental factors and $\tau_b$ is the bright exciton lifetime that can be measured or calibrated \cite{shilo_particle_2013,mazuz-harpaz_radiative_2016}. Fig. \ref{fig:summary}e shows the extracted value of $2n_b$ as a function of the excitation power. If there was a population equality between dark and bright particles ($n_d=n_b$), we would expect that $n_{tot}=n_d+n_b=2n_b$. This is certainly not the case at around $P_{c1}$, where we find $n_{tot}\gg n_b$, suggesting $n_d\gg n_b$ and a large imbalance between dark and bright particles. If indeed there is a large accumulation of dark excitons, which can only decay through tunneling to the contacts, we expect an increase in the measured current through the trap below $T_c$. This is indeed the case, as can be seen in Fig. \ref{fig:summary}h. In steady state the total loss rate (i.e. emitted photons+electrons in current) should stay constant under a constant generation rate. This is indeed verified in Fig. \ref{fig:summary}h, where we see that below $T_c$ the photon rate decreases to compensate for the increase in the tunneling rate, so that total loss stays the same. We note that as the excitation power increases above $P_{c1}$ the degree of darkening becomes smaller, until it disappears just before $P_{c2}$. While in the current results there is no direct experimental way to assign the  bright and dark populations to the corresponding total spin projection of excitons  $\pm1$ or $\pm2$, the connection to the theory put forward by \citet{combescot_``gray_2012} is quite compelling. Of course, at this point while we do not have any different picture to explain the above experimental findings, we cannot rule out that such alternative does exist. This is discussed in the next section.  

\begin{figure}
\centering
\includegraphics[width=0.5\textwidth]{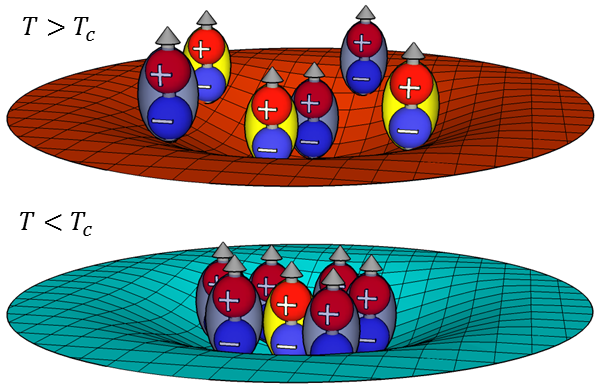}
\caption{\label{fig:gasLiquidIllustration} A schematic illustration of the observed states of an IX fluid trapped in a quasi-parabolic potential: A repulsive bright gas phase with equal populations of dark and bright IXs for $T>T_c$ and a dark, spatially compact, and dense liquid phase for $T<T_c$. See text for details. The bright and dark IXs are depicted by yellow and gray halos, respectively.}
\end{figure}

\section{Summary, discussion, and outlook}

We can conclude that a transition to a dense phase of dipoles is observed below a critical temperature $T_c$ and above a critical excitation power $P_{c1}$. This phase transition, schematically depicted in Fig. \ref{fig:gasLiquidIllustration}, is characterized by the following signatures:

\begin{enumerate}
\item \textbf{Short range order and close packing}: A correlated gas of IXs spontaneously increases its density as the temperature is dropped below $T\cong4.8K$, under a fixed optical excitation power. This density saturates at $T\cong2K$ to a high value $n_l\approx 6\mbox{-}9\times 10^{10}cm^{-2}$ satisfying $r_0n_l^{1/2}\gg1$, that is almost independent on the excitation power over at least two orders of magnitude. This sharp spontaneous increase of density is accompanied by a strong reduction of the spectral linewidth, indicating a phase transition to a dense phase with a short-range order and, importantly, with a well defined density that is spontaneously set by the system itself. Remarkably, this density range corresponds to approximately close packing of the IX wavefunctions $\expval{r_l}/a_b\sim 2\mbox{-}3$.

\item \textbf{Spatial condensation}: The spontaneous increase of density is accompanied by a strong shrinkage of the IX cloud area, indicating either a strong suppression of the IX mobility, a strong suppression of the thermal distribution of the kinetic energy, or the onset of attractive short-range interactions. Furthermore, a comparison of the experimental findings with a dipolar liquid model \cite{laikhtman_exciton_2009} indicates that the liquid phase is expected to be within the parameter range of a quantum liquid, having a significant quantum kinetic energy contribution and the quantum degeneracy condition $\lambda_{db}n_l^{1/2}\gg1$ should be well satisfied (see details in \citet{cohen_dark_2016}).

\item \textbf{Macroscopic accumulation of dark particles}: Below $T_c$ there is an evidence that the sharp buildup of density to the liquid state is driven mostly by accumulation of dark particles. At the low excitation onset of the transition to the liquid state, more than 90\% of the particles are in the dark state. This darkening is diminished as the excitation power is further increased, qualitatively similar to the prediction by \citet{combescot_``gray_2012}. This observed accumulation of dark particles might also explain how the sharp density buildup occurs above a tiny excitation power of $P\sim 10nW$. We also observed that the optical darkening of the IX fluid is quantitatively compensated by an increase of the tunneling current through the sample, which is the remaining depletion channel for dark particles that cannot recombine radiatively. 

\item \textbf{Well defined phase boundaries}: We found that the dark liquid phase at low temperatures is bounded by a gas phase from the low density side and by a bright, highly mobile plasma phase on the high density side. This is an initial partial mapping of the phase diagram of cold dipolar IX fluids. 

\end{enumerate}

All the above recent findings, by other research groups as well as by ours, opened up a narrow window to what seems as a complex phase diagram of cold dipolar IX fluids. As it seems, this phase diagram is the result of an intricate interplay between collective quantum effects, interaction driven correlations, and internal spin degrees of freedom, all bound together with the dynamics of a dissipative system with a finite lifetime. Remarkably, IX fluids might contain many-body physics even richer than initially thought. There are many fundamental open questions that are now awaiting an answer. Few of these questions are listed below:

\begin{enumerate}
\item Up until now, only a fraction of the full phase diagram of cold dipolar IX fluids have been mapped. Mapping of the full $(n,T)$ phase diagram is an outstanding challenge that is at the heart of understanding the system. Particularly interesting is to get the full mapping of the gas, liquid, and plasma phases, and of the dark liquid phase. An interesting open question is whether there are two distinct transitions, namely one to a classical liquid and one to a dark quantum liquid, or rather there is only one liquid state. Strikingly, the observed onset density $T_c$ in our experiment is in almost a perfect agreement with the critical temperature for the onset of the liquid phase reported in \citet{stern_exciton_2014}, even though the experiments were done with different samples and in different experimental geometries. That suggests that the observations might be strongly related and if they indeed are, it is interesting to understand what determines this specific temperature.  

\item The formation mechanisms and the nature of the ground state of the collective liquid states observed are far from being fully understood. In particular, several questions stand out: what drives the condensation transition in which the density spontaneously increases? Is it a classical phase transition, or is it driven by quantum degeneracy, as in the quantum liquid or BEC case? Is the phase transition purely thermodynamic, or is it driven dynamically by the unique coupling of the internal spin degrees of freedom of the IX to light? Our experiments indicate that dark IX spin states might play an important role, as a condensation to dark states which are long lived is at least consistent with many of our observations in Refs. \cite{shilo_particle_2013,cohen_dark_2016}, as well as in the work reported in \cite{beian_spectroscopic_2015}. More insight into the nature of the darkening and the connection to dark excitons is essential. 

\item More information on the collective ground state is certainly needed. A measurement of its compressibility, microscopic spatial correlations, momentum distribution, and excitation spectrum, are important to fully understand its nature. 

\item The issue of optical coherence also calls for additional study. As mentioned above, while several works reported the onset of extended coherence from a cold fluid of IXs, the works reporting the liquid formation were unable to measure spatial coherence beyond the diffraction limit. One possible explanation is that the coherence is hidden in the dark part of the fluid in those experiments. Another option is the size of the dipole. As is shown in Ref. \cite{lozovik_strong_2007}, the single particle density matrix gets strongly localized for increasing dipole size, which is expected to reduce the spatial coherence significantly. Indeed, in the works that reported extended optical coherence IXs with smaller dipoles were studied compared to those reporting a liquid formation. Finally, if the liquid state is purely classical, no coherence is expected, but then the question arises why this liquid forms in some experiments and not in others. A consistent experimental mapping of the effect of the dipole size has not yet been done.

\end{enumerate}

Hopefully, future experiments and more comprehensive theoreis will soon elucidate the open questions that remain to be answered. Whatever these answers would eventually be, it is clear that they will involve interesting physics of interacting particles.

\bibliographystyle{model1-num-names}
\bibliography{My_Library.bib}

\end{document}